\documentclass[jcp,twocolumn,superscriptaddress,floatfix,showpacs]{revtex4-1}  
\usepackage{graphicx}
\usepackage{hyperref}

\begin{document}

\author{Benedict E. K. Snodin}
\affiliation{Physical \& Theoretical Chemistry Laboratory, Department of Chemistry, South Parks Road, Oxford, OX1 3QZ, United Kingdom}
\author{John S. Schreck}
\affiliation{Department of Chemical Engineering, Columbia University, 500 W 120th Street, New York, NY
10027, USA}
\author{Flavio Romano}
\affiliation{Dipartimento di Scienze Molecolari e Nanosistemi, Universit Ca' Foscari, Via Torino 155, 30172 Venezia Mestre,
Italy}
\author{Ard A. Louis}
\affiliation{Rudolf Peierls Centre for Theoretical Physics, University of Oxford, 1 Keble Road, Oxford, OX1 3NP,
United Kingdom}
\author{Jonathan P. K. Doye}
\affiliation{Physical \& Theoretical Chemistry Laboratory, Department of Chemistry, South Parks Road, Oxford, OX1 3QZ, United Kingdom}

\title{Coarse-grained modelling of the structural properties of DNA origami}

\begin{abstract}
We use the oxDNA coarse-grained model to provide a detailed 
characterization of the fundamental structural properties of DNA origamis, 
focussing on archetypal 2D and 3D origamis.
The model reproduces well the characteristic pattern of helix bending in 
a 2D origami, showing that it stems from the intrinsic tendency
of anti-parallel four-way junctions to splay apart, a tendency that is enhanced
both by less screened electrostatic interactions and by increased thermal
motion.  
We also compare to the structure of a 3D origami whose structure has been
determined by cryo-electron microscopy. The oxDNA average structure has a
root-mean-square deviation from the experimental structure of 8.4\,\AA, which is
of the order of the experimental resolution. These results illustrate that the
oxDNA model is capable of providing detailed and accurate insights into the structure
of DNA origamis, and has the potential to be used to routinely pre-screen 
putative origami designs.
\end{abstract}

\maketitle

\section{Introduction}
DNA nanotechnology seeks to use the specificity of the Watson-Crick base
pairing and the programmability possible through the DNA sequence to design
self-assembling nanoscale DNA structures and devices. The most prevalent
technique used is probably that of DNA origami in which a long viral
``scaffold'' DNA single strand is folded up into virtually any arbitrary
structure by the addition of many different ``staple'' strands that bind to
multiple specific domains on the scaffold \cite{Linko13,Hong17}. The initial
designs were two-dimensional \cite{Rothemund06} but were soon generalized to
three-dimensional shapes \cite{Douglas09}, and then to bent,
twisted \cite{Dietz09} and curved \cite{Han11} structures through the
introduction of internal mechanical stresses. The increasing usage of origamis
was particularly facilitated by the development of computer-aided design tools,
such as cadnano \cite{Douglas09b}. These original approaches produced structures
involving mainly bundles of locally parallel double helices held together by
four-way junctions. More recently, scaffolded origami approaches have been
developed that generate more open ``wireframe''
structures \cite{Benson15,Veneziano16,Matthies16}.
The structural control and the addressability provided by the DNA origami technique naturally have led to 
many types of applications, particularly in the areas of biosensing, drug delivery and nanofabrication \cite{Wang17,Liu18}. 

In Rothemund's original paper the structure of the origamis were characterized
by atomic force microscopy (AFM). The images were used to confirm that the
origamis had folded into the designed structures without significant defects,
and identified structural features of the origamis, such as what we here term
the ``weave'' 
pattern where the helices, rather than being straight, splay out between 
four-way junctions, thus leading to the characteristic pattern where the
helices weave back and forth between adjacent helices \cite{Rothemund06}. Such
microscopy studies (by AFM and transmission electron microscopy) are probably
the most prevalent way of characterizing the structures of DNA origamis, but
are usually limited in terms of the fine-grained detail that can be obtained.
Furthermore, adsorption onto a surface may perturb the structure, especially
for 2D origamis, which may be flattened and made to look more ordered because
of the suppression of out-of-plane thermal fluctuations. 

Solutions-based measurements can be performed by, for example, small-angle
X-ray scattering (SAXS) and FRET, but SAXS interpretation usually requires a
structural model (and its computed SAXS pattern) for
comparison \cite{Andersen09,Fischer16,Bruetzel16,Baker18}. FRET can potentially provide
detailed measurements of selected distances, but has been relatively little
used to provide detailed structural analysis of origamis \cite{Stein11,Funke16}.

Cryo-EM can potentially provide the most detailed structural analysis. For
example, Bai {\it et al.} were able to obtain a high-resolution structure for a
three-dimensional 
origami where an all-atom structure was fitted to
the obtained electron density maps \cite{Bai12}. However, such detailed studies
are unlikely to be a routine approach. More commonly, cryoEM has been used at a
lower level of resolution, particularly for polyhedral nanostructures \cite{He08,Kato09}.
Very recently, particle electron tomography has also begun to be applied 
allowing visualization of the 3D structure of individual 
DNA nanostructures \cite{Lei18,Wagenbauer17b}.

Given both the difficulty of obtaining high-resolution structural information
and the potential utility of being able to predict structural properties prior
to experimental realization, computational modelling of the structure of DNA
origamis has the potential to play a significant role in the
field \cite{Jabbari15}. All-atom simulations have the potential to provide the
most detailed structural
insights \cite{Yoo13,Wu13,Li15,Gopfrich16,Maffeo16,Lee17}. Notably, the
Aksimentiev group have simulated a number of
origamis \cite{Yoo13,Li15,Gopfrich16,Maffeo16},
including even an origami nanopore inserted into a membrane \cite{Gopfrich16}.
However, such simulations are extremely computationally intensive and cannot be
performed routinely. 
Furthermore, even for the relatively stiff origamis considered in these
studies, it is not clear that they have fully equilibrated on the simulation
time scales \cite{Maffeo16}. More promising as a general tool is an approach
where only the atoms of the origami (but not the water environment) are
simulated and an elastic network is used to constrain the origami in its
assembled state; these constraints are applied to the base pairing and base
stacking interactions, and also to the distance between neighboring
helices \cite{Maffeo16}. 

A computationally less expensive approach is to use coarse-grained models in
which the basic units are no longer atoms, but some larger moiety, be it a
nucleotide \cite{Ouldridge11,Sulc12,Snodin15}, a base
pair \cite{mergell03,Arbona12} or a section of a double
helix \cite{Castro11,Kim12,Pan14,Sedeh16,Reshetnikov18,Hemmig18}. Such
approaches of course inevitably lead to a lower level of structural detail, and
the accuracy of their properties will depend on the quality of the
parameterization. 

By far the most widely-used approach is ``cando'' as it allows efficient and
reliable structural screening of potential origami designs through a
simple-to-use web interface \cite{Castro11,Kim12,Pan14,Sedeh16}. However, its
lack of excluded volume interaction means that it may not be appropriate for
flexible origamis whose structure is not fully mechanically constrained.
Furthermore, as with any model whose basic unit is above the level of a
nucleotide, there is no coupling to intra-base-pair degrees of freedom;
consequently processes such as duplex fraying, junction migration, and breaking
of base pairs due to internal stresses cannot be resolved. Finally, it has a 
simplified representation of single-stranded DNA, and so cannot take into
account, for example, secondary structure formation.

All these potential deficiencies can be addressed by a
nucleotide-level model, albeit at greater computational expense.  Although
there are a number of such models at this level of
detail \cite{MorrisAndrews10,Hinckley13,Chakraborty18}, here we explore in
detail the description of DNA origamis provided by the oxDNA
model \cite{Ouldridge11,Sulc12,Snodin15}. This model has been particularly
successful at describing a wide variety of biophysical properties of
DNA \cite{Ouldridge11,Romano13,Ouldridge13b,Matek15,Snodin15,Harrison15,Skoruppa17},
and has been applied to a significant number of DNA nanotechnology
systems \cite{Ouldridge10,Ouldridge13,Doye13,Sulc14,Machinek14,Kocar16,Snodin15,Snodin16,Schreck16,Shi17,Sharma17,Khara18,Fonseca18,Engel18}.

What are the features that make the oxDNA model particularly appropriate to
study DNA origamis? Firstly, it is able to accurately reproduce DNA's basic
structural properties. Properties such as the DNA pitch are particularly
important, as the large size of DNA origamis means that small deviations can
lead to internal stresses that lead to global twisting of the origami---note
that in the second version of the oxDNA model the duplex pitch and the twist at
nicks and junction were fine-tuned to correct just such an
issue \cite{Snodin15}. Secondly, it is able to capture the mechanical
properties of DNA such as the persistence length and torsional
modulus \cite{Ouldridge11,Matek15,Snodin15,Skoruppa17}; these are 
important for correctly capturing both the thermal fluctuations of DNA origami
and the equilibrium structure when internal stresses are deliberately designed
into the origami to cause overall bend and twist \cite{Dietz09}. Thirdly, it has
a very good representation of the thermodynamics of
hybridization \cite{Ouldridge11,Sulc12,Snodin15} allowing it to capture fraying,
the breaking of base pairs due to more extreme internal or external stresses,
and secondary structure formation in single strands. Fourthly, it has a good
representation of the mechanical properties of single-stranded
DNA \cite{Sulc12}; this is relevant to that subset of origamis that use ssDNA to
introduce flexibility \cite{Marras15}, exert forces \cite{Nickels16b} or brace
tensegrity structures \cite{Liedl10}.

OxDNA is also able to naturally capture the mechanical behaviour of other
sites. For example, in unstressed DNA it is generally favourable for the DNA to
stack across a nick and in this state it has very similar elastic properties to
standard duplex DNA. However, for a relatively small free-energy cost this
stacking can be broken and the two halves of the duplex can then rotate
relatively freely about the hinge point \cite{Harrison15}. Lower-resolution
models
are typically unable to capture the two-state character associated with the
nick and instead associate a single set of moduli with the nick.

OxDNA has previously been used to characterize a number of specific DNA
origamis with good success \cite{Snodin15,Shi17,Sharma17,Khara18}. Here, our aim
is to provide a detailed structural analysis of some of the basic features of
DNA origami and to test the reliability of the oxDNA model by comparing to the
most structurally detailed available experimental data. The systems that we
will study are an archetypal two-dimensional origami tile that has recently been 
characterized by SAXS \cite{Baker18}, and the 3D
origami of Ref.\ \cite{Bai12}. To better understand the origins of some
of the structural features, we also characterize the free-energy landscape of
an unconstrained four-way junction, a key motif in DNA origamis.

\section{Materials and Methods}

In this work, we have used the version of the oxDNA model described in Ref.\
\cite{Snodin15} (sometimes called ``oxDNA2'') for which the properties have
been fine-tuned to capture origami twist. This version of the model also has an
explicit dependence on the ionic strength through a Debye-H\"{u}ckel-like term
in the potential.  Note, such a simple form is, of course, not capable of
capturing ion-specific effects. This electrostatic term was parameterized so
that the dependence of duplex melting on [Na$^+$] is reproduced. However, oxDNA
overestimates the stability of the stacked form of the Holliday junction as a
function of [Na$^+$]. As it happens, this is an advantage when modelling DNA
origamis, as in experiments origamis are typically assembled in a buffer that
contains Mg$^{2+}$, which is known to stabilize the stacked form of the
Holliday junction in an ion-specific manner \cite{Lilley00}. Thus,
[Na$^+$]=0.5\,M, the solution conditions at which we chose to model the
origamis using oxDNA, is a reasonable choice to mimic the experimental solution
conditions, with only very small changes in the structural properties of the
oxDNA origamis occurring as the concentration is further increased. By contrast,
extremely high values of [Na$^+$] have to be used in experiment to induce
origami assembly \cite{Martin14} probably because of the relative instability of
the stacked Holliday junction in [Na$^+$] solutions \cite{Lilley00}. We also
note that, we used the oxDNA model with average-strength (rather than
sequence-dependent) interactions, as we are interested in generic structural
properties. 

To generate initial origami structures, we have developed a publicly available
script to turn a cadnano file into a starting oxDNA configuration.
The initial structures generated may locally be subject to very large forces
due to overlaps or somewhat extended backbones. As such large forces are 
potentially problematic for a simulation,  
we have a developed an algorithm to first 
relax configurations prior to them being simulated.

The molecular dynamics simulations use a strongly coupled thermostat
to generate diffusive motion of the nucleotides in the absence of explicit 
solvent. To aid the simulation of large origamis, we use a GPU-enabled verion
of our simulation code \cite{Rovigatti15}. 

\begin{figure*}
\includegraphics[width=7in]{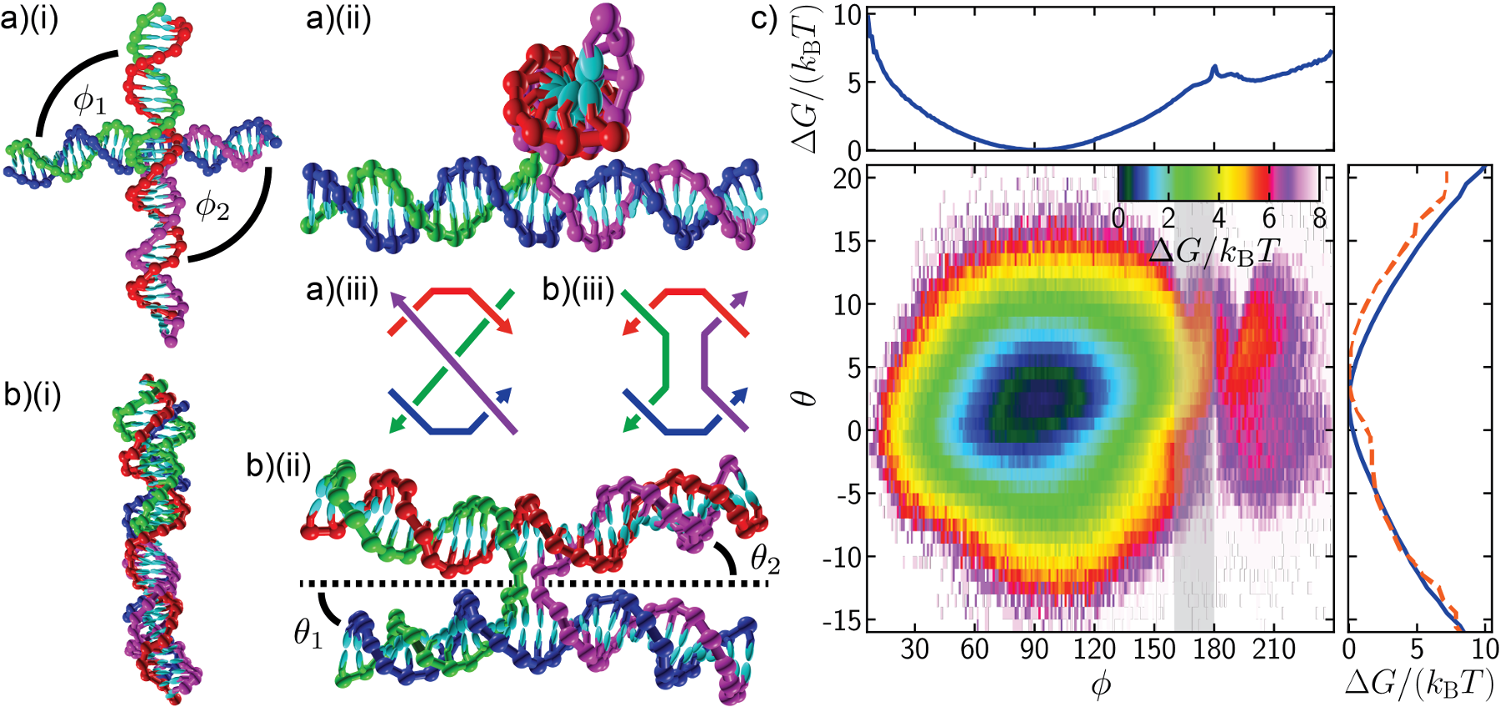}
\caption{(a) and (b) OxDNA configurations for a single Holliday junction at
0.5\,M salt that are representative of (a) the left-handed global free-energy
minimum ($\phi=95.5^\circ,\theta=2.5^\circ$) and (b) an anti-parallel junction
($\phi=180^\circ,\theta=5^\circ$). (i) and (ii) provide perpendicular views of
the configurations.  In both cases the green and purple strands cross from one
double helix to another, while the blue and red strands carry straight on along
their double helix. (a)(iii) and (b)(iii) provide schematics of a left-handed
parallel and an anti-parallel junction, respectively. 
(c) The Holliday-junction free-energy landscape as a function of $\phi$ and
$\theta$. One-dimensional free-energy profiles for $\phi$ and $\theta$ are also
shown.  The full free-energy profile for $\theta$ is shown as a solid blue
line, while the dashed red line is for the subregion $160^\circ \leq \phi \leq
180^\circ$, shaded grey in the free-energy landscape. 
$\phi=(\phi_1+\phi_2)/2$ is a measure of the chiral twist of the junction,
where $\phi_1$ and $\phi_2$ are the angles between the arms as indicated in
(a)(i).  $\theta=(\theta_1+\theta_2)/2$ is the average angle between each arm
and the plane of the junction, as indicated in (b)(ii).  
The temperature used was 296.15\,K.
}
\label{fig:junction}
\end{figure*}

\section{Results and Discussion}
\subsection{Holliday junction structure}

Before we directly address origami structure, we first consider the properties
of a single four-way junction, also known as a Holliday junction, as they are
an essential feature of origami designs. These occur wherever two strands,
usually staple strands,
cross from one double helix to another within the origami. Thus the junctions
play the vital role of joining adjacent double helices together. Because each
origami contains many such junctions, their structural properties can potentially
have a major effect on the structure of the origami.

The structure of a single isolated Holliday junctions has been characterised
experimentally \cite{Lilley00,Lilley09} through X-ray crystallography
\cite{Ortiz-Lombardia99,Eichman00}, AFM \cite{Mao99} and FRET measurements
\cite{Eis93,Hohng07}. Depending on the experimental conditions, a Holliday
junction can exist in an open or stacked conformation, with the open
conformation favoured at low concentrations of metal ions, and with divalent
ions being particularly effective at stabilizing the stacked conformation
\cite{Lilley09}. Because the adoption of a stacked geometry is important to the
structural integrity of standard DNA origamis, ``high salt'' conditions are
used in origami experiments, normally, but not necessarily \cite{Martin14}, in
a solution containing significant Mg$^{2+}$.  For this reason, here we only
consider the stacked conformation of the junction, this form being favoured by
oxDNA for the salt conditions we consider here (0.5\,M).

In order to quantify the structure of the junction, we define two angles:
$\phi$, which measures the average twist angle between pairs of arms; and
$\theta$, which measures the average angle between the arms and the plane of
the junction (Fig.~\ref{fig:junction}). $\phi$ is defined such that
$\phi=0^\circ$ for a parallel junction and $\phi=180^\circ$ for an
anti-parallel junction, where the (anti-)parallel character refers to the
relative orientation of the two strands that do not cross in the stacked
junction. The junctions in origamis are typically anti-parallel. A junction is
said to be right-handed if $\phi>180^\circ$ and left-handed if
$\phi<180^\circ$.  $\theta$ provides a measure of how much the DNA helices bend
at the junction and will be particularly relevant when we consider the
``weave'' pattern for DNA origamis.

\label{sec:free junction characterisation}

To characterize the free-energy landscape of the junction, we ran molecular 
dynamics simulations of an isolated Holliday junction with arms that are 16 base pairs long.
We used a biasing potential (defined in the Supplementary Data)
to window the simulations in $\phi$, with the particular aim of 
accelerating the sampling for $\phi$ values near $180^\circ$, as 
this is the region relevant to the junctions in origamis. The sequence was chosen to
prevent branch migration.  The junction can adopt one of two stacked isomers.
In order to simplify the analysis, we only considered one of the isomers and so
configurations that were determined to be in the wrong isomer were discarded.
Further details regarding the simulations, including the DNA strand sequences
and how a configuration's isomeric state is determined, are given in the
Supplementary Data.

The resulting free-energy landscape is depicted in Fig.~\ref{fig:junction}(c).
The free-energy minimum for the junction is at
$(\phi,\theta)=(95.5^\circ,2.5^\circ)$, while integrating over $\theta$ gives a
preferred $\phi$ angle of 90.5$^\circ$, with a mean value of 92.0$^\circ$. 
Thus, Holliday junctions prefer to be left-handed in the oxDNA model. However, the junctions
observed in crystal structures are usually right-handed with $\phi\approx 240^\circ$ \cite{Lilley09}. 
The preferred value of $\phi$ that
we see with oxDNA is simply that required to align the backbone sites of 
the two double helices at the junction. In this way the distance between the
two double helices can be maximized, reducing the steric and electrostatic repulsion
between them. At other angles the bonds between the two helices are twisted, bringing
the helices closer together at the junction.

The cause of this difference in the preferred geometry is probably the model's
simplified representation of the backbone and its associated excluded volume.
However, we note that we know of no coarse-grained DNA model
for which a right-handed junction naturally emerges from the model. For
example, the next most widely-used nucleotide-level coarse-grained model also
exhibits left-handed junctions \cite{Wang16}.
Furthermore, we also note that a crystallized left-handed junction has been
reported \cite{Nowakowski00} for an RNA-DNA complex, and that both chiral forms
have been seen as local minima for a junction in solution in all-atom
simulations \cite{Yu04}.

That oxDNA is unable to reproduce the experimental junction crystal structures'
preference to be right-handed is, fortunately, not particularly detrimental to
modelling origami structure with oxDNA for the following reasons. Firstly, the
helices are able to rotate relatively freely about the crossover point and so
the junction is relatively flexible in $\phi$ 
(although clearly the junction will very rarely adopt a
configuration with $\phi \sim 240^\circ$). 
Secondly, in an origami the junctions are constrained to $\phi$ values close to
$180^\circ$, i.e.~90$^\circ$ from the preferred junction angle for oxDNA and
60$^\circ$ for real DNA, so the junctions in both cases would be expected to
have a somewhat similar level of stress. Indeed, recent oxDNA results on a 2D
DNA brick system have shown that oxDNA can reproduce the melting point of such
systems very well (to within $2^\circ$C of experiment) without any
adjustment to the model \cite{Fonseca18}. This would suggest that oxDNA can
capture well the thermodynamic cost of the anti-parallel junctions in these DNA
brick systems, albeit noting that the junctions are less restricted than those
considered here as they involve only one strand crossing. 

Finally, junctions that have been experimentally resolved within a 3D origami
have been found to exhibit $\phi$ angles slightly below
$180^\circ$ \cite{Bai12}, a somewhat counter-intuitive result because one would
perhaps expect deviations to take the junction towards (not away from) the
preferred geometry. However, this is beneficial for oxDNA modelling of
origamis, because, as we will see, the origami junctions in oxDNA do somewhat
twist towards their preferred left-handed orientation. 
This tendency to have a slight left-handed twist has also been observed in
atomistic simulations of 3D origamis \cite{Yoo13}. Although the cause of this
left-handed twist in the experimental case is not obvious, it must reflect
both the details of the real free-energy landscape (as a function of $\phi$)
and the constraints placed on the junctions within the origami.  For oxDNA,
there is a maximum at $\phi \sim 180^\circ$, caused by the larger repulsions
when the two helices are aligned, but with a greater slope away from the
maximum towards the preferred left-handed configurations. 

The free-energy profile for $\theta$ in Fig.~\ref{fig:junction}(c)
shows a slight preference for a positive $\theta$, with a free-energy minimum
at $\theta=2.5^\circ$, corresponding to a tendency for the helix arms to bend
slightly away from the plane of the junction. The effect is greater, with the
minimum at $\theta=4.5^\circ$, for a subset of the data for which $160^\circ
\leq \phi \leq 180^\circ$, the region likely to be relevant within origamis. 
A simple argument explaining this behaviour 
is that negative values of $\theta$ will cause the arms to bump into
each other more often, and this becomes more likely when the arms are
approximately aligned, as for $\phi \approx 180^\circ$. 
Furthermore, the twisting of the inter-helix bonds for anti-parallel junctions
will lead to increased repulsion local to the junction.

Intriguingly, however, this argument would lead us to expect a similar effect
as we move towards $\phi \approx 0^\circ$, but this effect is not evident from
the free-energy landscape; instead the landscape becomes more symmetric about
$\theta=0$ in this region. Interestingly, the free-energy cost of forming a
parallel junction is also much more than that of an anti-parallel junction.
This may help to explain why DNA nanostructures with parallel junctions that
reliably assemble have generally been more difficult to
design \cite{Kumara08,Han13b}.
As $\phi \approx 0^\circ$ configurations are not relevant to junctions within
conventional DNA origamis we do not investigate the origins of these effects
further.

\subsection{Structural properties of a 2D origami}
\label{sec:2D structure}

The coupling between many Holliday junctions present in a DNA origami generates
a rich set of structural properties. We first consider 2D origamis, which
consist of a single ``sheet'' of (anti-)parallel DNA helices joined by
crossovers.  The particular design on which we focus 
has a very regular pattern of crossovers and staples (see Fig.~S2
for the cadnano representation of the structure). Rothemund's
original 2D origami tiles have been shown to be somewhat twisted \cite{Li12b},
because of a slight mismatch between the pitch of DNA (about 10.5 base pairs)
and the separation between the junctions (32 base pairs for three helical
turns).  The current design has included a suitable number of 
sections with 31 base pairs between equivalent junctions to compensate for this
effect.

The average structure of the origami is shown in Fig.~\ref{fig:2Dtile} 
(see Supplementary Data 
for details of how the average structure was computed). 
The origami sheet is not noticeably twisted, but there is some modest
curvature (as also predicted by cando).  
SAXS experiments on this tile by Baker {\it et al.} are consistent with a
flat to moderately curved shape \cite{Baker18}.
Although the structure does fluctuate considerably, here
we focus on the average structure. Note that, when adsorbed onto a surface (as
is the case for most experimental structural studies of origamis), rather than
in solution, we would expect the structure to be flattened out and much less
fluxional.

\begin{figure*}
\begin{centering}
\includegraphics[width=7in]{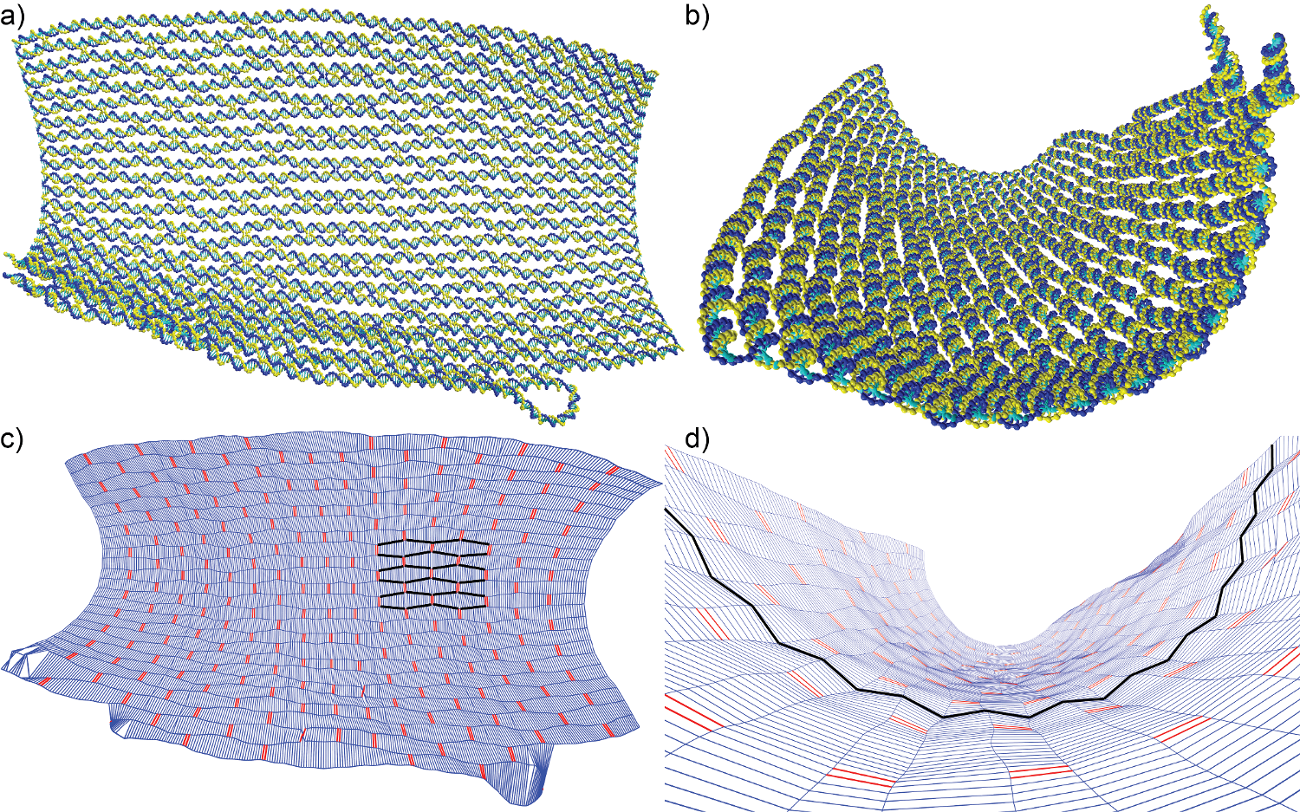}
\caption{Different representations and viewpoints of the average structure of the 2D origami.
(c) and (d) provide a ``chickenwire'' representation of the tile. In (c) the tile is shown from the front
so that the weave pattern can be clearly seen (as highlighted by helix axes in black); and 
(d) at an angle to show the corrugation pattern on the tile's surface (as highlighted by
the inter-helix vectors in black). In (c) lines running horizontally along the origami show the axes of the double helices that make up the origami; pairs of red vertical lines represent double crossovers, while blue vertical lines represent the inter-helix vectors used for the quantitative analysis of the weave and corrugation. 
Deviations from the typical structure, such as that seen in the bottom left corner of the origami, 
are caused by staple melting or branch migration.}
\label{fig:2Dtile}
\end{centering}
\end{figure*}

The ``weave pattern'' in 2D origamis, where adjacent double helices tend to
push apart and open up a significant gap between the helices away from the
junctions, has been well known since the DNA origami method was originally
devised, and is clearly visible in experimental microscopy
images \cite{Rothemund06}. It is also very apparent in the average structure 
depicted in Fig.~\ref{fig:2Dtile}. In Rothemund's original paper he suggested
two possible reasons for this behaviour: firstly, electrostatic repulsion
between the negatively charged helices; and secondly, that the detailed local
structure around the junctions favours the helical arms to bend slightly away
from each other \cite{Rothemund06}.  
A third possible contribution is an entropic effect due to the increased
conformational space available when adjacent double helices are not perfectly
parallel.  Here, we will see that all three of these effects play a role in
the origin of the weave pattern for 2D origamis in our oxDNA simulations.

\begin{figure}
\begin{centering}
\includegraphics[width=3.3in]{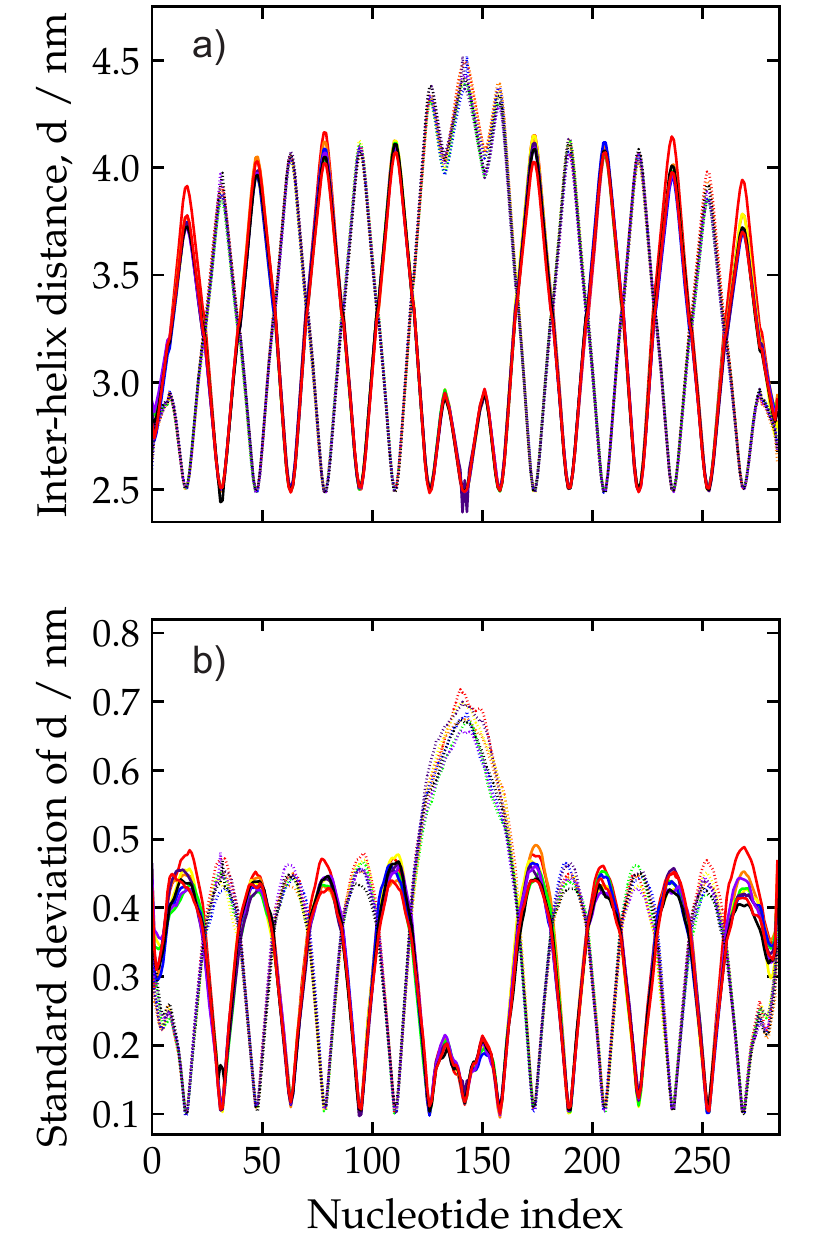}
\caption{The weave pattern for the 2D tile at 300\,K and $[\text{Na}^+]=0.5\,\text{M}$, quantified by (a) the inter-helix distance and (b) the standard deviation in this distance as a function of base-pair index along the origami. Each line corresponds to a different pair of adjacent double helices on the origami. Some pairs have been omitted for clarity (see main text). The symmetry of the design is such that the pairs may be split into two groups: here one group is plotted with solid lines and one with dotted lines.
}
\label{fig:weave plot}
\end{centering}
\end{figure}

We quantify the weave pattern of the 2D tile by measuring the distance between
the helix axes (defined as the midpoint between the bases for each base pair)
for adjacent double helices. 
The results shown for the tile at a temperature of 300\,K 
and $[\text{Na}^+]=0.5\,\text{M}$ are plotted in Fig.~\ref{fig:weave plot}.
(Note, to simplify the appearance of the plot we omit some inter-helix
distances.  Namely, those involving the double helices at the top and bottom
edges of the origami, as these are only constrained on one side and so exhibit
slightly different behaviour; a few affected by branch migration which resulted
in spurious results near the affected junction; and one affected by a partially
melted staple that caused enhanced flexibility.) Because of the regular pattern
of junction placement in the origami's design (Fig.~S2),
there are two obvious groups into which the pairs of double helices can be
divided. 
Each group exhibits a wave-like pattern with minima at the crossovers, where
the double helices are brought closest together, and maxima away from the
crossovers, normally at a position which is both midway between the junctions
and where the adjacent pair of helices have a crossover. This pattern has a
periodicity of about 32 base-pair steps, corresponding to the periodic junction
placement in the origami. 

In the middle of the plot (around base-pair index 150), a different pattern is
evident. This is due to the presence of the origami's seam (a series of
junctions where the scaffold strand is exchanged),
which runs along the middle of the tile. In this region, one group of
double-helix pairs has a particularly large section without any junctions and
so opens up to the largest extent heree\cite{Ma18}, as is also very clear from
Fig.~\ref{fig:2Dtile}; the modulations in the distance in the middle of this
region reflect the presence of junctions on adjacent pairs of helices.  By
contrast, the other group of double-helix pairs has a shorter distance between
junctions due to the extra scaffold crossovers, and opens up much less.

It is also interesting to note the ``triangular wave'' character of the weave
plots. The bending that creates the weave pattern is mostly localized at the
junctions with the intervening sections basically straight. This is in part
because the junctions are relatively flexible compared to the duplex sections
between the junctions, which being only a small fraction of the persistence
length (typically only 16 base pairs long compared to the 125 base pairs for
the duplex persistence length for the model \cite{Snodin15}) are very stiff.
Furthermore, we can examine the fluctuations in the weave pattern, quantified
in Fig.~\ref{fig:weave plot}(b) by the standard deviation in the inter-helix
distances. The plot shows that the fluctuations, which are smallest at the
junctions and largest at the midpoints between the junctions, are
significantly smaller in magnitude than the variation in the interhelical
distance due to the weave pattern itself. Thus, the junctions in the origami
are very unlikely to adopt a configuration where the helices are straight with
no weave (i.e.\ a value of $\theta$ near to zero) and have a static structural 
preference to be bent away from each other. 

This is consistent with the picture that we obtained from the single Holliday
junctions free-energy landscape, where for junctions that were near to
anti-parallel, the free energy as a function of $\theta$ had a minimum at
significantly positive $\theta$ (e.g.\ $4.5^\circ$ for junctions in the range
$160^\circ<\theta< 180^\circ)$.
We can also estimate the preferred value of $\theta$ at a junction
in an origami from the weave pattern. Assuming a perfectly triangular wave form
and taking 1.5\,nm as a typical value of the difference in the interhelix
distance between the maxima and minima of the weave pattern (Fig.\
\ref{fig:weave plot}(a)) together with their 16 base-pair separation gives
$\theta=7.85^\circ$.

\begin{figure*}
\centering
\includegraphics[width=7in]{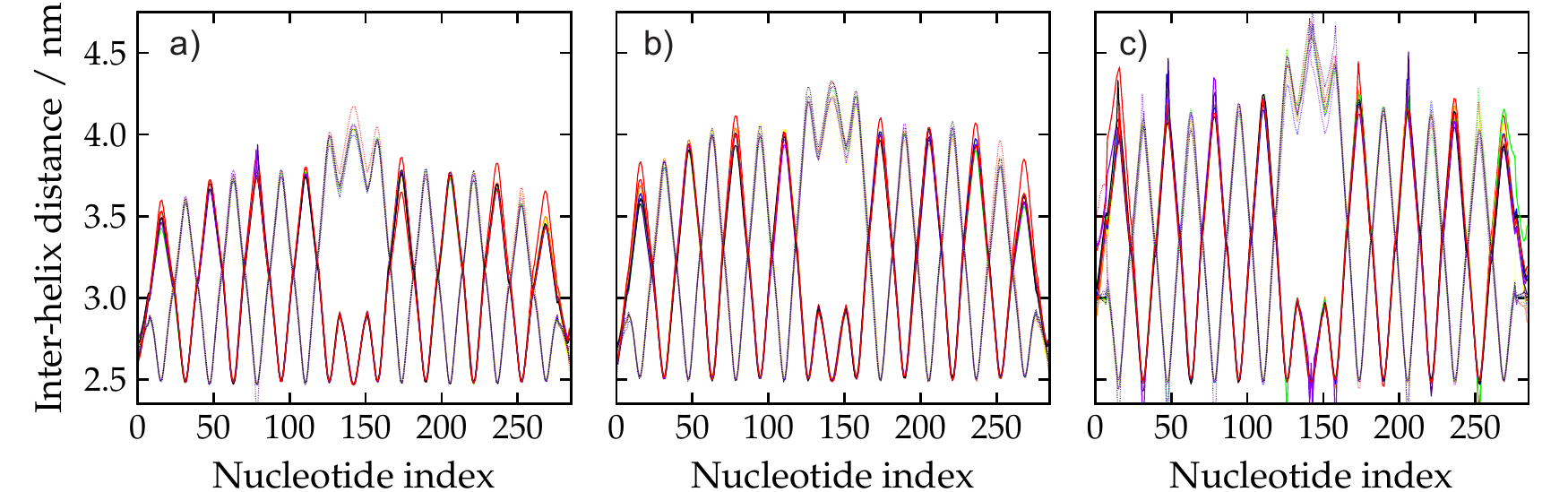}
\caption{The weave pattern for the 2D tile (a) when the electrostatic term in the model is removed, and 
at $[\text{Na}^+]=0.5\,\text{M}$ and a temperature of (b) 270\,K and (c) 330\,K. 
The data is presented in an analogous way to Fig.~\ref{fig:weave plot}.}
\label{fig:weave other}
\end{figure*}

In order to investigate the effect of electrostatic repulsion, we repeated the
simulations of the 2D tile with the electrostatic term in the potential
removed.
The result is shown in Fig.~\ref{fig:weave other}(a). We found that the weave
pattern remained, albeit with a reduction in the magnitude of the oscillations
by about 20\%. 
This indicates that, although electrostatic repulsion enhances the weave
pattern in oxDNA, it is not the sole cause.  Experimental evidence for an
increase in the inter-helical spacing for 3D origami as the ionic strength is
decreased has been recently observed by SAXS of a 24-helix
bundle \cite{Fischer16} and by cryoEM of covalently-cross-linked
origami \cite{Gerling18}.

Finally, we simulated the tile at different temperatures 
to obtain further insight into whether the weave pattern has a partly entropic,
as well as energetic, origin. Fig.~\ref{fig:weave other}(b) and (c) show the
weave pattern at 270\,K and 330\,K. Together with the weave pattern at 300\,K
(Fig.~\ref{fig:weave plot}), the plots indicate that the magnitude of the
oscillations characterising the weave pattern increases somewhat with increasing
temperature. Thus, thermal fluctuations play a role in determining the
magnitude of the weave pattern, with this entropic component favouring a more
pronounced weave pattern. Again, this is consistent with the free-energy
landscape in Fig.\ \ref{fig:junction}(c) where the asymmetry of the minimum in the
effective free-energy profile as a function of $\theta$ for anti-parallel
junctions suggests that the average $\theta$ should increase as thermal
fluctuations increase.

A second structural property that we see in 2D origamis is what we term
``corrugation'', where the origami displays a systematic, out-of-plane bending of
the double helices, so that the junctions have a $\phi$ angle that is not
exactly 180$^\circ$, as would be the case for an origami with perfectly
anti-parallel double helices. Due to the regularity of the origami design, this results 
in a wave-like pattern, albeit much smaller in magnitude than the weave pattern, on the surface of the origami that 
is visible for average-structure configurations, as shown in Fig.~\ref{fig:2Dtile}(d). 
That the free-energy for a free Holliday junction is a maximum for a perfectly 
anti-parallel junctions provides the driving force for this effect; for oxDNA
this leads to a left-handed twist to the junctions.

Our approach to measure the corrugation is to follow how the orientations of
the inter-helix vectors vary as one moves away from a junction.  Specifically,
for every such inter-helix vector we measure the angle between the inter-helix
vector and the average inter-helix vector at the nearest junction between that
pair of helices when projected onto the plane perpendicular to the average
helix axis at that junction. The sign of the angle is determined from the sign
of the scalar triple product of the two projected inter-helix vectors with the
average helix axis, in order to distinguish clockwise and anticlockwise
twisting. Thus, this measure quantifies the amount of twisting, in the plane
perpendicular to the helix axis, between adjacent double helices near to
junctions.

\begin{figure}
\begin{centering}
\includegraphics[width=3.3in]{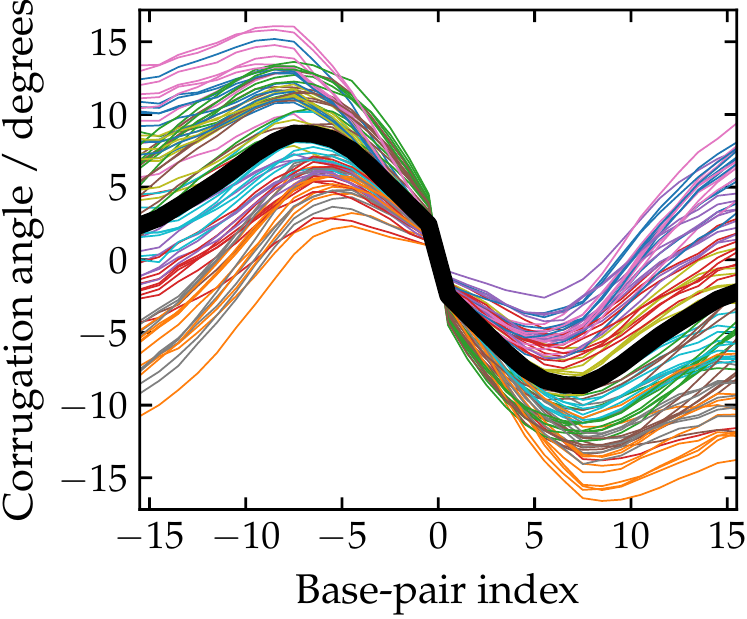}
\caption{The corrugation pattern for the 2D tile. The angle between inter-helix vectors is is shown on the $y$-axis (see the main text for details of the definition). The $x$-axis shows the location of the inter-helix vector, in base-pair steps relative to the midpoint of the junction. Each line corresponds to a different junction, with the thick black line being an average over the data. Some junctions have been omitted for clarity (see main text).}
\label{fig:corrugation plot}
\end{centering}
\end{figure}

The corrugation for the 2D tile as measured with this method is plotted in
Fig.~\ref{fig:corrugation plot}. Note, we do not include data for all
junctions, but only those that have the canonical pattern of neighbouring
junctions; thus, we exclude the outermost junctions on the tile, and the
junctions next to the scaffold seam as well as the seam itself. The plot shows
the tendency for the double helices to come slightly out of the plane of the
tile. The interhelix vectors are systematically rotated in one direction on one side of the junction (base
pair index less than 0) and in the opposite direction on the other (base pair index
greater than 1). This corresponds to a $\phi$ angle of less than 180$^\circ$
for each junction, which is as expected from the properties of the free
junction. 

Although each junction shows qualitatively the same behaviour, there
is clearly a wide variation in the curves for the base 
pairs furthest from the junction. Much of this
is due to the method of twist compensation used in the origami design.
Although overall the origami is roughly untwisted, this is achieved by 
having a mixture of interjunction double-helical sections that are over-
and under-twisted (with 31 and 32 base pairs between junctions, respectively).

The general shape of the curves can be understood by considering the pattern of
junctions in the origami.
In order that adjacent junctions between a pair of double helices have the same twist, the
curves must have (at least) one complete waveform every 31/32 base pairs. Thus,
at the midpoint between the junctions the chiral twist angle must pass through
zero. This position corresponds to the positions of junctions between
adjacent helices and allows them also to have a left-handed chiral twist. 
Note that, unlike the weave pattern, the corrugation requires the 
maximum bending in between rather than at junctions, and so will be resisted
by the bending stiffness of the duplex.

\begin{figure*}
\centering
\includegraphics[width=7in]{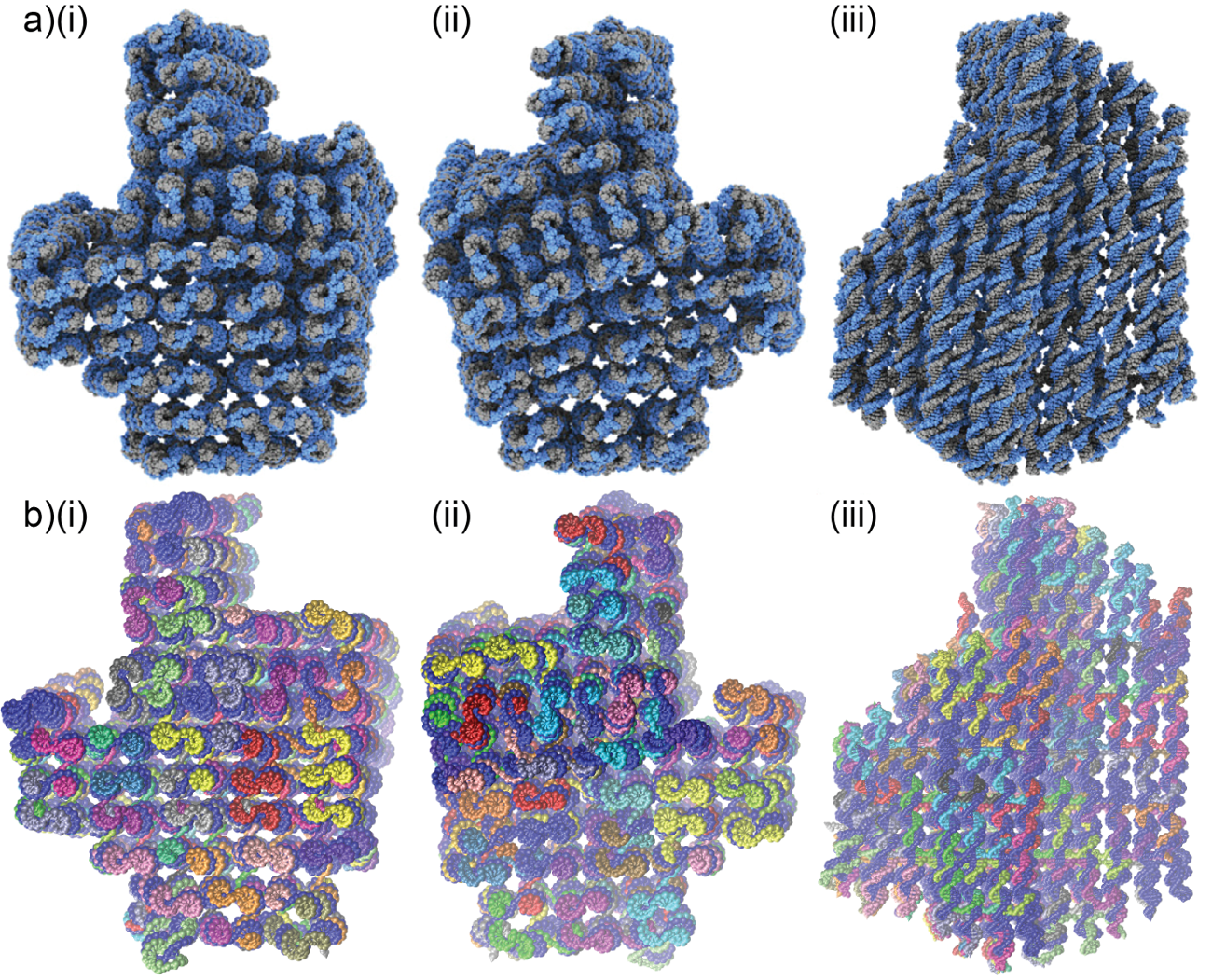}
\caption{Three different views of the 3D origami structure determined by (a) fitting an atomistic model to cryo-EM data 
(reproduced 
from Ref.~\cite{Bai12})
and 
(b) computing the average structure in simulations with the oxDNA model.}
\label{fig:dietz pointer comparison}
\end{figure*}

Although, to our knowledge, this corrugation effect has not been reported in
any experimental studies of 2D origamis, this is perhaps not surprising because
the effect is small in magnitude and would tend to be reduced or removed when
the structure is placed on a surface to be visualised, as is usually the case
in experiments. 
The underlying cause of the corrugation is the tendency of the junctions to
twist away from the perfectly anti-parallel conformation. 
Out-of-plane ``distortions'' have also been noted in a series of planar
``ring'' origamis when modelled using cando \cite{Pan14}, although in this case
the junctions have a right-handed twist because the experimental Holliday
junction conformation was set as the minimum of the harmonic junction twist
potential.

\subsection{Structure of a 3D origami}

In this section, we further test the ability of oxDNA to accurately model DNA
origami by making a detailed comparison to the 3D origami of Ref.\ \cite{Bai12} 
whose structure was characterised in detail by cryo-EM. The design uses a square
lattice for the double helices, which means that the crossovers are spaced 32
base pair steps apart,
and we expect a slight global twist in the structure.
The average structure computed from oxDNA simulations (see Supplementary Data
for details of the averaging
procedure) is compared to the experimentally-determined structure in
Fig.~\ref{fig:dietz pointer comparison}. 
By eye, the two structures appear very similar.

\begin{figure*}
\centering
\includegraphics[width=7in]{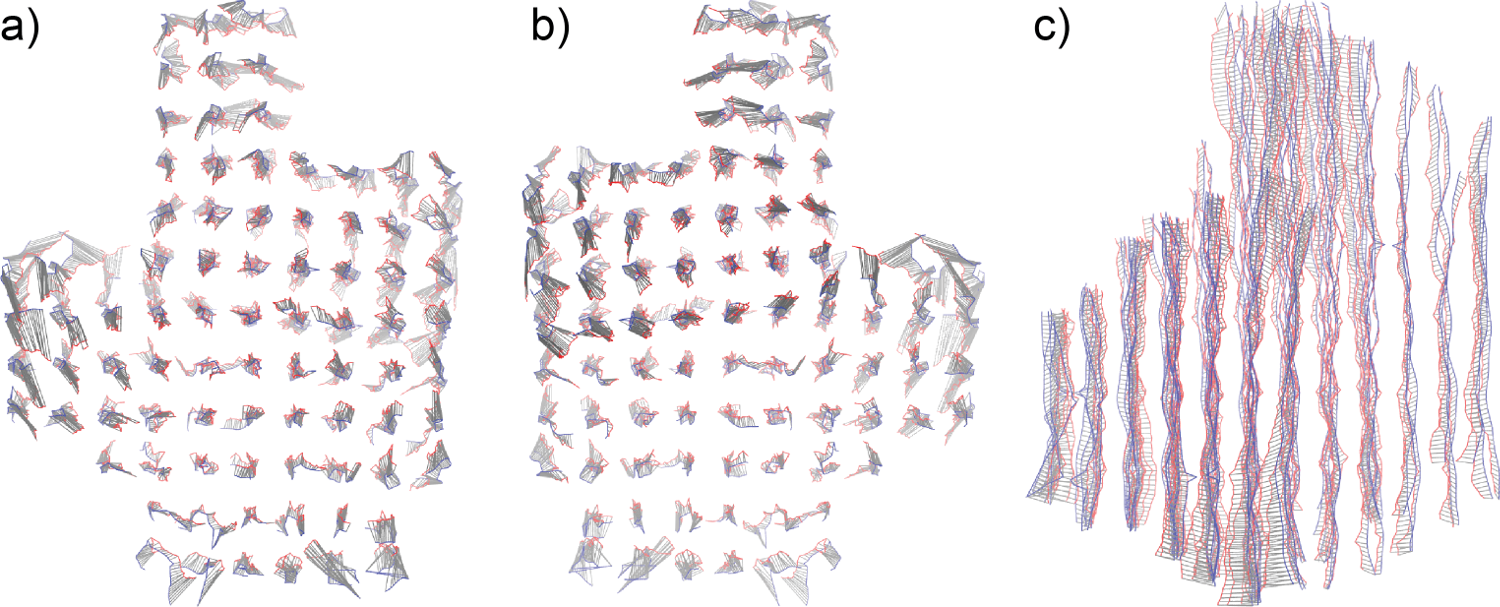}
\caption{The aligned helix axes of the 3D origami that were used for the RMSD calculation (views as in Fig.\ \ref{fig:dietz pointer comparison}).  The simulated structure is in blue, the experimental one is in red, and the grey lines show the displacement vectors.} 
\label{fig:dietz pointer comparison bbms}
\end{figure*}

To quantify this further we calculate the square root of the mean squared
displacement (RMSD) between the simulated structure and the experimentally
determined one. (Details of the RMSD calculation are given in the Supplementary Data
.) We find that the RMSD is 0.84\,nm, an
excellent agreement with experiment. A graphical comparison is shown in
Fig.~\ref{fig:dietz pointer comparison bbms}. From this figure, it is clear that the
overall size of the cross-sectional lattice predicted by oxDNA is very close
to the experimentally determined one, indicating that the magnitude of the weave
pattern and the radius of the DNA double helix match experiment well. In addition, the
overall twist of the structure is reproduced. The majority of the contribution
to the RMSD is due to the double helices towards the outside of the structure,
which are more clearly displaced from the experimentally determined structure.
One potential reason for this disparity is that our average structure includes
the effects of thermal fluctuations at room temperature, and it is not clear to
what extent these fluctuations will be frozen in during the cryo-EM process. We
also note that the estimated resolution of the cryo-EM characterization of the
origami is reported as 0.97\,nm at the core and 1.4\,nm at the periphery,
comparable to the RMSD we have found. 
For further comparison, fully-atomistic simulations of this origami were able
to achieve an RMSD of 1.1\,nm, which improved in the elastic-network guided
simulations to 0.9\,nm \cite{Maffeo16}.

\begin{figure*}
\centering
\includegraphics[width=7in]{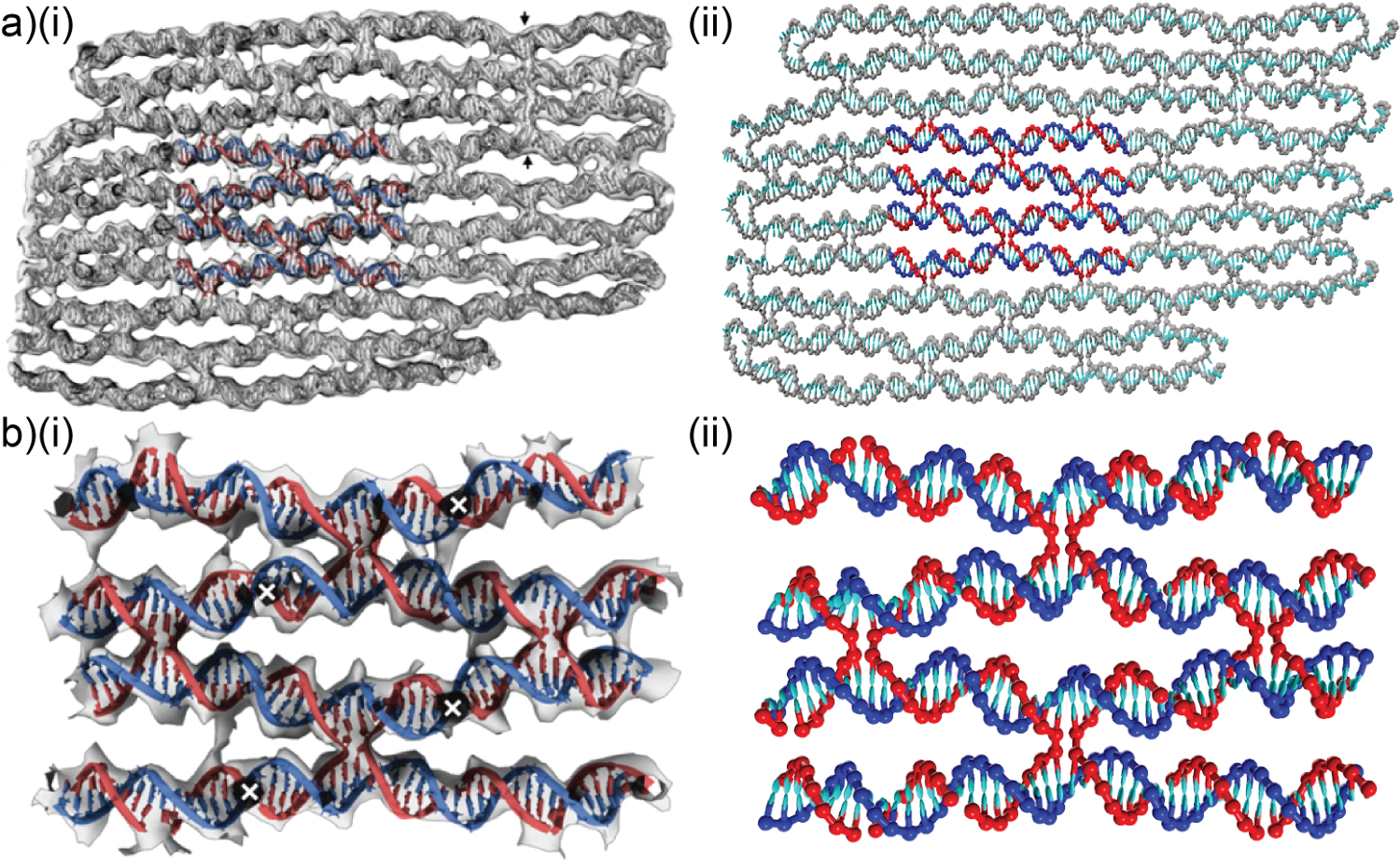}
\caption{A comparison between experimental and oxDNA images for a typical plane from the 3D origami.
(a) shows the whole plane and (b) a close-up of the highlighted region in (a).
Images in (a)(i) and (b)(i) are reproduced from Ref.\ \cite{Bai12}.
}
\label{fig:dietz pointer comparison detail}
\end{figure*}

\begin{figure*}
\centering
\includegraphics[width=7in]{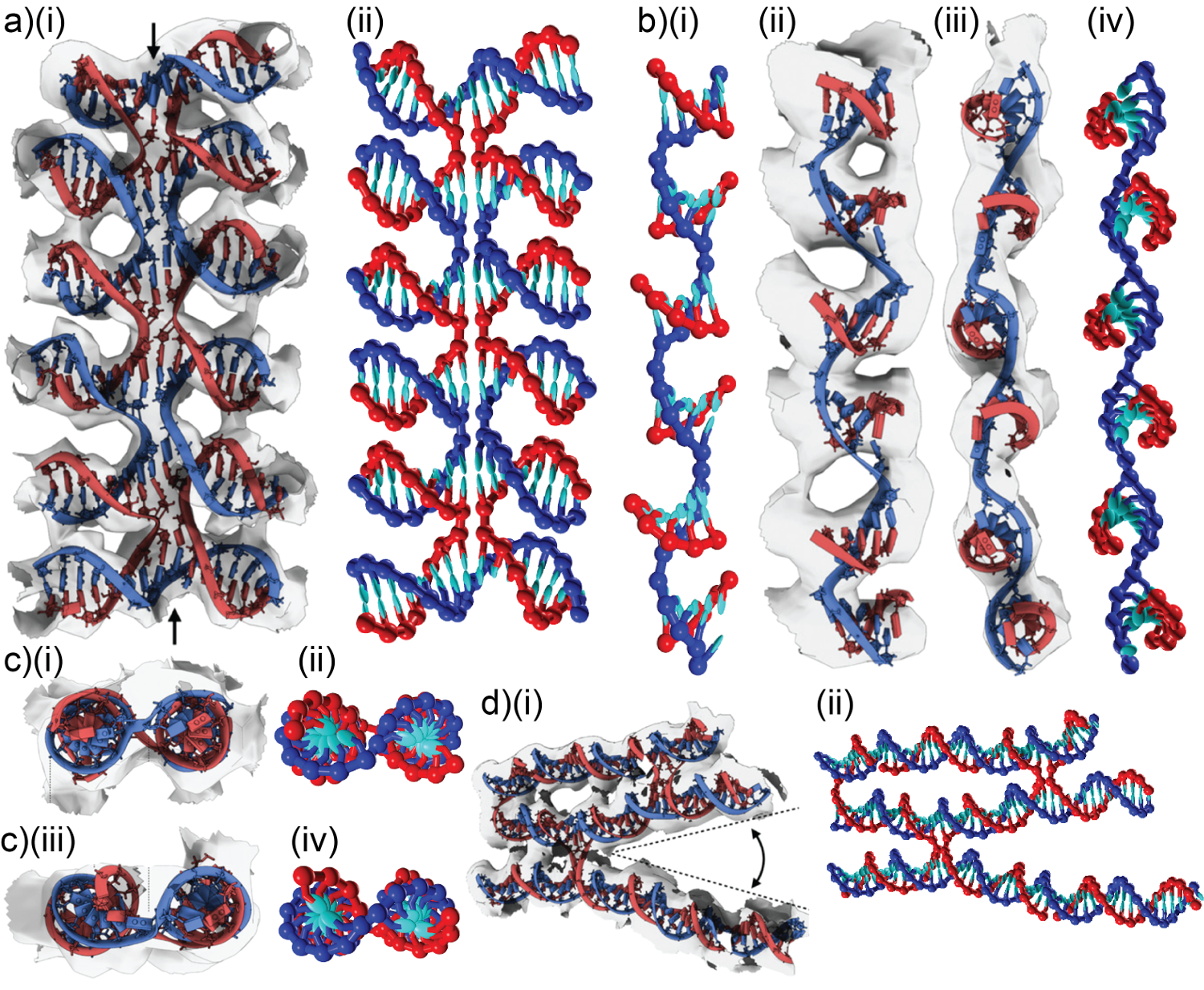}
\caption{A comparison between experimental and oxDNA images for a variety of detailed motifs from the 3D origami.
(a) Five aligned Holliday junctions that are held in place by coaxial stacking,
(b) A section where the scaffold adopts a left-handed pseudo-helix associated with a set of five adjacent short domains.
(c) A comparison of (i) and (ii) a typical Holliday junctions in the origami with (iii) and (iv) one that is under twist stress
due to there being one fewer base pair in an adjacent helix (often called a ``deletion'').
(d) shows an example of the greater splaying of helices that can occur at the ends of an origami due to a lack of constraints 
from other junctions.
Images in (a)(i), (b)(ii) and (iii), (c)(i) and (iii), and (d)(i) are reproduced from Ref.\ \cite{Bai12}.
}
\label{fig:dietz pointer comparison detail2}
\end{figure*}

Further evidence for the reliability of the oxDNA structure can be obtained by
visual comparison of a variety of motifs from the origami.  The examples, 
depicted in Figures \ref{fig:dietz pointer comparison detail} 
and \ref{fig:dietz pointer comparison detail2}, are the same as those chosen in
Ref.\ \cite{Bai12}.  The similarity between the experimental and simulation
images is very apparent. 
In particular, Fig.\ \ref{fig:dietz pointer comparison detail} depicts a plane
from the origami that, for the most part, has a similar pattern of junctions to the
2D origami, and therefore shows a very similar weave pattern, 
whereas Fig.\ \ref{fig:dietz pointer comparison detail2} provides a comparison of some more
detailed motifs.

\begin{figure*}
\centering
\includegraphics[width=7in]{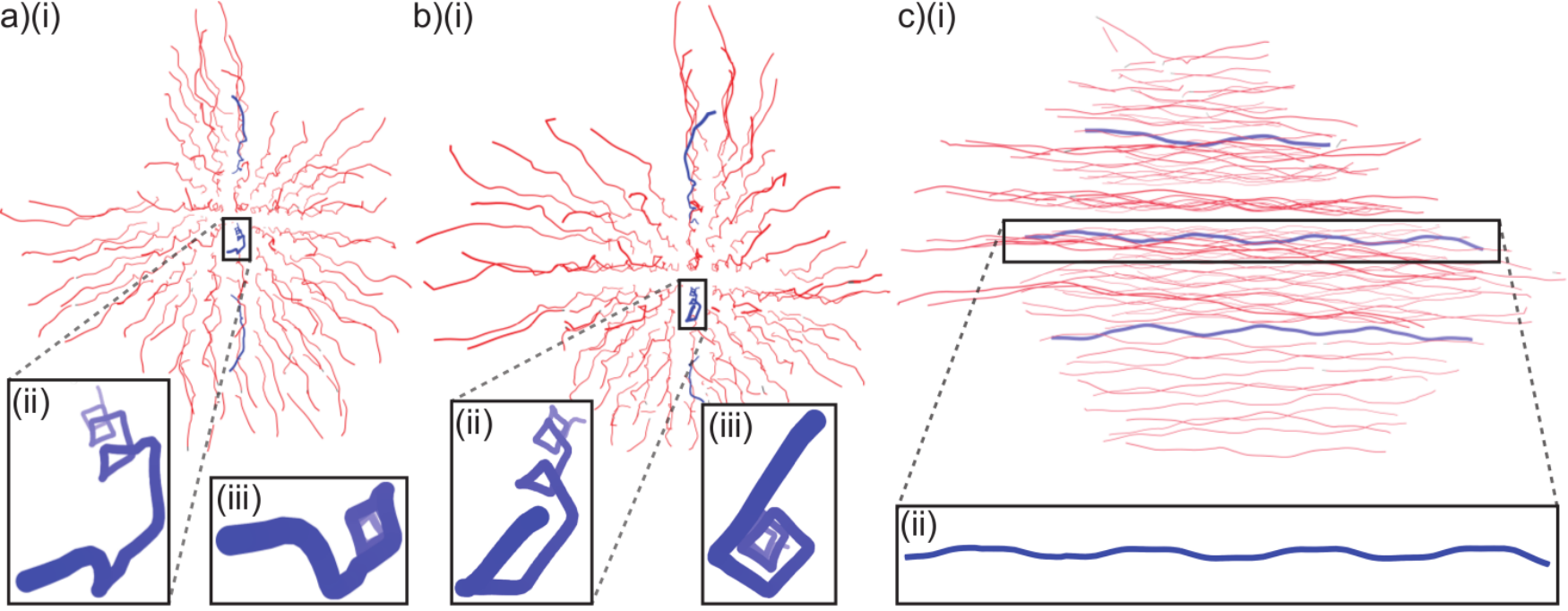}
\caption{A representation of the 3D origami, where the red lines show the 
paths of the centres of the individual double helices from three different viewpoints. 
The pattern of junctions leads to these paths having a weak left-handed helicity, 
as highlighted by the blue paths in the insets.}
\label{fig:helical}
\end{figure*}

One intriguing feature that we have identified is that the 3D origami's
double-helix axes trace out a left-handed helix with a period of approximately
32 base-pair steps per turn, which corresponds to the spacing of junctions
between each adjacent double helix pair in this design.  This can be seen in Fig.\ \ref{fig:helical}.
We note that this weak feature is not clear in the experimental data perhaps because 
of the greater noise in the data.

This effect is mainly due to
the ``weave'' distortion, because unlike in the case of the 2D origami, the
junctions do not lie in a single plane but in two orthogonal planes. In
particular,
as one moves along a double helix, it is drawn closer to each of its four
neighbouring double helices in turn, as each junction is encountered. 
The junctions trace a left-handed helical path because they are separated 
by three-quarters of a pitch length along the right-handed double-helical sections. 
This path is also consistent with a left-handed chiral twist to each individual junction. 
Finally, that the helices possess a very weak toroidal writhe may be relevant
when considering the precise effective pitch needed to generate an untwisted origami.

\section{Conclusion}

In this paper, we have used the oxDNA coarse-grained model to characterize in
detail some of the fundamental structural features of 2D and 3D DNA origami, 
and made a detailed comparison to the most detailed experimentally determined
origami structure. In particular, we have shown that the weave
pattern associated with the splaying of helices at the junctions has its origin
in the basic structural properties of anti-parallel Holliday junctions, and is 
further enhanced by electrostatic repulsion and thermal fluctuations.
For 2D origamis, we have also found a weaker out-of-plane corrugation
associated with the slight left-handed twist of the Holliday junctions in origamis.

Our comparison to the 3D origami of Ref.\ \cite{Bai12} confirms the suitability of 
the oxDNA model for structural characterization of DNA origami. 
Structures that have been carefully characterised experimentally will give further
opportunities to test and refine the structural predictions of the model, while
for DNA origamis that have only been visualised using low-resolution methods,
oxDNA has the potential to provide more detailed structural insights, as has
already been done for a number of examples \cite{Shi17,Sharma17}.
The model could also be used to pre-screen the properties of putative origami
designs prior to experimental realization to aid the design process.

The capabilities of the oxDNA model should be seen as complementary to other
computational strategies for origami structure prediction at different levels
of detail. The particular advantages over more coarse-grained approaches
potentially include explicit representations of excluded volume and base
stacking, and realistic descriptions of single-stranded DNA and the breaking of
base pairs, albeit, of course, at a greater computational cost. 
Furthermore, oxDNA provides a well-tested model for which a wide-range of
biophysical properties of DNA are known to be be accurately described.
Examples of where these features will be particularly useful include origamis
with flexible components \cite{Sharma17}, origamis under significant internal
stresses, and origamis where single-stranded components play a key functional
role.  The model also allows the loss of origami structure under thermal or
external stresses \cite{Engel18} to be investigated.
However, if a more atomically detailed view of origamis than is available
through oxDNA is required, all-atom approaches are likely to be most
appropriate.



\section{Acknowledgements}
We acknowledge the use of the University of Oxford Advanced Research Computing
facility [doi:10.5281/zenodo.22558]. JSS acknowledges computing resources from
Columbia University's Shared Research Computing Facility project.

\section{Funding}
This work was supported by the Engineering and Physical Sciences Research Council 
(Grants EP/I001352/1 and EP/J019445/1).  


%

\end{document}